\documentclass{PoS}

\title{Open Charm decays and spectroscopy at Belle}

\ShortTitle{Open Charm decays \& spectroscopy}

\author{\speaker{B. Pal, A. Schwartz}\thanks{On behalf of the Belle Collaboration}\\
        University of Cincinnati\\
        E-mail: \email{palbs@ucmail.uc.edu}}

\abstract{In this review we report the recent results of open charm decays and  spectroscopy  using  the  data  collected  with  the  Belle  detector  at the KEKB asymmetric-energy
$e^+e^-$ collider.
  }

\FullConference{VIII  International Workshop on Charm Physics\\
		5-9 September, 2016\\
		Bologna, ITALY}

\begin{document}

\section{Introduction}
In this report an overview of the recent results of open charm decays and  spectroscopy is presented based on the  data, collected by the Belle experiment at the KEKB $e^+e^-$ asymmetric-energy collider~\cite{KEKB}. 
(Throughout this paper charge-conjugate modes are implied.) 
The experiment
took data at center-of-mass energies corresponding
to several $\Upsilon(nS)$ resonances; the total data sample
recorded exceeds $1~{\rm ab}^{-1}$.

The Belle detector is a large-solid-angle magnetic
spectrometer that consists of a silicon vertex detector
(SVD), a 50-layer central drift chamber (CDC),
an array of aerogel threshold Cherenkov counters
(ACC), a barrel-like arrangement of time-of-flight
scintillation counters (TOF), and an electromagnetic
calorimeter comprised of CsI(Tl) crystals
(ECL) located inside a super-conducting solenoid
coil that provides a 1.5 T magnetic field. An iron
flux-return located outside of the coil is instrumented
to detect $K^0_L$
mesons and to identify muons
(KLM). The detector is described in detail elsewhere~\cite{Belle, svd2}.

%%%%%%%%%%%%%%%%%%%%%%%%%%%%%%%%%%%%%%%%%%%%%%%%%%%%%%
\section{Analysis of $D^{**}$ production}
Orbitally excited states of the $D$ meson ($D^{**}$ states) provide a good opportunity to test the heavy quark effective theory (HQET)~\cite{Neubert:1993mb} and QCD sum rule~\cite{Uraltsev:2000ce} predictions.
Precise knowledge of the properties of the $D^{**}$ state is important to reduce uncertainties in the measurements of the semileptonic decays and thus in the determination
of the Cabibbo-Kobayashi-Maskawa matrix elements $|V_{cb}|$ and $|V_{ub}|$. 
Many $D^{**}$ mesons have already been observed by B factories and LHCb~\cite{Agashe:2014kda}.
In this analysis, we perform  an amplitude analysis of the $\bar{B}^0\to D^{*+}\omega\pi^-$ decay to measure the decay fractions of $D^{**}$ states and to study the $D^{**}$ properties~\cite{Matvienko:2015gqa}, using $711~{\rm fb}^{-1}$ of data collected at the $\Upsilon(4S)$ resonance.

The total signal yield is obtained from a binned $\chi^2$ fit to the $\Delta E=\sqrt{|\vec{p}|^2+m^2}-E_{\rm beam}$ distribution, with a tight cut in the beam-constrained mass $M_{\rm bc}$. The corresponding branching fraction is measured to be
$$
\mathcal{B}(\bar{B}^0\to D^{*+}\omega\pi^-)=(2.31\pm0.11\pm0.14)\times10^{-3}.
$$
Unless explicitly stated otherwise, whenever two uncertainties are quoted in this proceeding, the first is statistical and the second is systematic.
The result is consistent with the values of CLEO~\cite{Alexander:2001fp} and BaBar~\cite{Aubert:2006zb} but with higher precision.

\begin{figure}[htb!]
\begin{center}
\includegraphics[width=\textwidth]{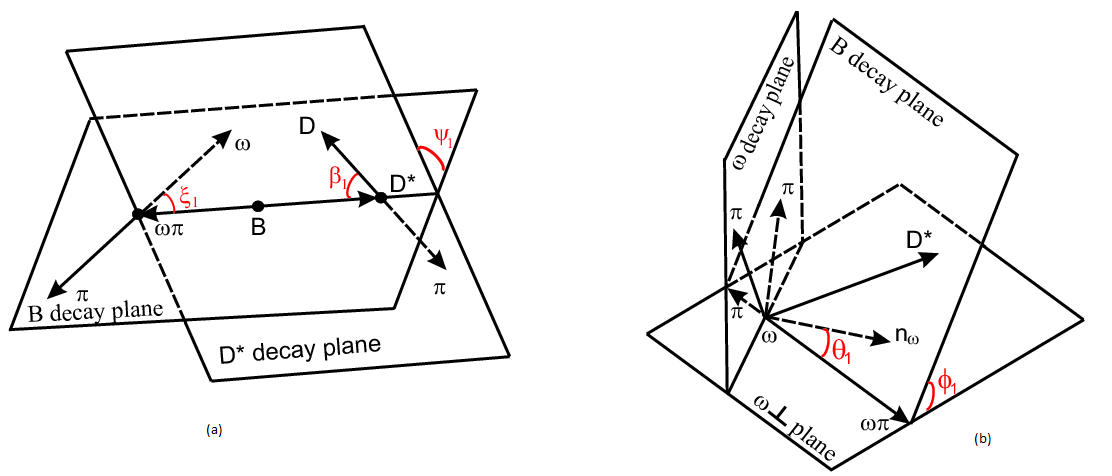}
\caption{\small Kinematics of a $\bar{B}^0\to D^{*+}\omega\pi^-$ decay mediated by an $\omega\pi^-$ intermediate resonance. The diagram in (a) defines two polar angles $\xi_1$ and $\beta_1$ and one azimuthal angle $\psi_1$. The diagram in (b) defines one polar angle $\theta_1$ and one azimuthal angle $\phi_1$. The direction $n_{\omega}$ in (b) corresponds to the vector normal to the $\omega$ decay plane.}
\label{fig:ang_var}
\end{center}
\end{figure}

A six-dimensional amplitude analysis is performed using the method described in Ref.~\cite{Matvienko:2011ic}. We define two sets of variables: [$M^2(\omega\pi)$, $\cos\theta_1$, $\phi_1$, $\cos\beta_1$, $\psi_1$,  $\cos\xi_1$] and  [$M^2(D^*\pi)$, $\cos\theta_2$, $\phi_2$, $\cos\beta_2$, $\psi_2$, $\cos\xi_2$], corresponding to the $\omega\pi$ and $D^{**}$ production, respectively. The masses $M(\omega\pi)$ and $M(D^*\pi)$ are the invariant masses of the $\omega\pi$ and $D^*\pi$ combinations. The angular variables 
describing $\omega\pi$ production are defined in Fig.~\ref{fig:ang_var}. 
The angular variables describing $D^{**}$ production are defined in the same manner as angles for $\omega\pi$ production, but with the $D^*\pi$ flight direction instead of the $\omega\pi$.

\begin{figure}[htb!]
\begin{center}
\includegraphics[width=\textwidth]{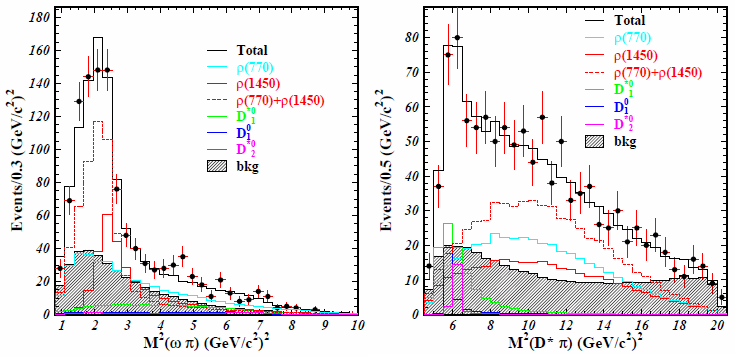}
\caption{\small Distributions of the variables (left) $M^2(\omega\pi)$ and (right) $M^2(D^*\pi)$. }
\label{fig:dalitz}
\end{center}
\end{figure}

To describe all the features of the Dalitz plot, we use the following set of resonances: off-shell $\rho(770)^-$, $\rho(1450)^-$, $D_1(2430)^0$, $D_1(2420)^0$, $D_2^*(2460)^0$ and $b_1(1235)^-$. Figure~\ref{fig:dalitz} shows the projections of $M^2(\omega\pi)$ and $M^2(D^*\pi)$ distributions and the results are summarized in Table~\ref{tab:dalitz}.

\begin{table}[htb]
\renewcommand{\arraystretch}{1.2}
\caption{\small Summary of the final results of the  $\bar{B}^0\to D^{*+}\omega\pi^-$  amplitude analysis. The first error is statistical, the second is systematic and the third is the model error.}
\label{tab:dalitz}
\centering
\begin{tabular}{l|c}\hline
\hline
Parameter & Result \\
\hline
$\mathcal{B}(\bar{B^0}\to\rho(770)^-D^{*+})$ & $(1.48\pm0.27^{+0.15+0.21}_{-0.09-0.56})\times10^{-3}$\\
\hline
$\mathcal{B}(\bar{B^0}\to\rho(1450)^-D^{*+})$ & $(1.07^{+0.15+0.06+0.40}_{-0.31-0.13-0.02})\times10^{-3}$\\
Mass of $\rho(1450)^-$ & $(1544\pm22^{+11+1}_{-~1-46})$ MeV/$c^2$\\ 
Width of $\rho(1450)^-$ & $(303^{+31+3+69}_{-52-4-~6})$ MeV \\
\hline
$\mathcal{B}(\bar{B^0}\to D_1(2430)^0\omega)$ & $(2.5\pm0.4^{+0.7+0.4}_{-0.2-0.1})\times10^{-4}$ \\
S-wave fraction & $(38.9\pm10.8^{+4.3+1.2}_{-0.7-1.1})$\%\\
P-wave fraction & $(33.1\pm9.5^{+2.4+3.0}_{-5.5-4.0})$\%\\
D-wave fraction & $(28.3\pm8.9^{+3.0+3.9}_{-0.8-2.9})$\%\\
Longitudinal polarization & $(63.0\pm9.1\pm4.6^{+4.6}_{-3.9})$\%\\
\hline
$\mathcal{B}(\bar{B^0}\to D_1(2420)^0\omega)$ & $(0.7\pm0.2^{+0.1}_{-0.0}\pm0.1)\times10^{-4}$ \\
Longitudinal polarization & $(67.1\pm11.7^{+0.0+2.3}_{-4.2-2.8})$\%\\
\hline
$\mathcal{B}(\bar{B^0}\to D_2^*(2460)^0\omega)$ & $(0.4\pm0.1^{+0.0}_{-0.1}\pm0.1)\times10^{-4}$ \\
Longitudinal polarization & $(76.0^{+18.3}_{-8.5}\pm2.0^{+2.9}_{-2.0})$\%\\
\hline
$\mathcal{B}(\bar{B^0}\to b_1(1235)^-D^{*+})$ & $<0.7\times10^{-4}$ at 90\% confidence level \\
\hline
\hline
\end{tabular}
\end{table}
%%%%%%%%%%%%%%%%%%%%%%%%%%%%%%%%%%%%%%%%%%%%%%%%%%%%%%
%======================================================================
\section{Analysis of charmed baryon decays}
%\subsection{Analysis of the decay $\Lambda_c^+\to\phi p\pi^0$}

%%%%%%%%%%%%%%%%%%%%%%%%%%%%%%%%%%%%%%%%%%%%%%%%%%%%%%
\subsection{First observation of the doubly Cabibbo-suppressed $\Lambda_c^+$ decay }
Several doubly Cabibbo-suppressed (DCS) decays of charmed mesons have been observed~\cite{Agashe:2014kda}. Their measured branching ratios with respect to the corresponding Cabibbo-favored (CF) decays play an important role in
constraining models of the decay of charmed hadrons and in the study of flavor- $SU(3)$ symmetry~\cite{Lipkin:2002za, Gao:2006nb}. On the other hand, because of the smaller production cross sections for charmed baryons, DCS decays of charmed
baryons have not yet been observed. Here we present the first observation of the DCS decay $\Lambda_c^+\to pK^+\pi^-$ and the measurement of its branching ratio with respect to the CF decay $\Lambda_c^+\to pK^-\pi^+$, using $980~{\rm fb}^{-1}$ of data~\cite{Yang:2015ytm}.
\begin{figure}[htb!]
\begin{center}
\includegraphics[width=\textwidth, height=6.5cm]{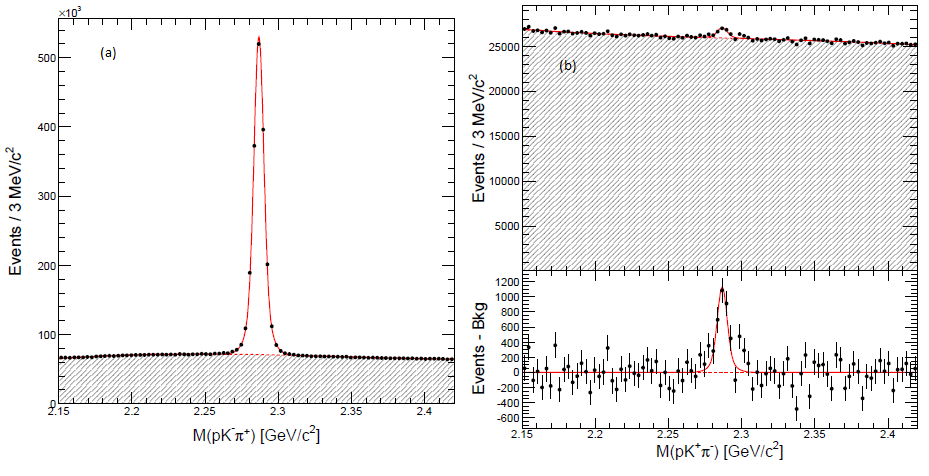}
\vskip -0.3cm
\caption{\small Distributions of (a) $M(pK^-\pi^+)$ and (b) $M(pK^+\pi^-)$ and residuals of data with respect to the fitted combinatorial background. The solid curves indicate the full fit model  and the dashed curves the combinatorial background.}
\label{fig:dcs}
\end{center}
\end{figure}

Figure~\ref{fig:dcs} shows invariant mass distributions of (a) $pK^-\pi^+$ (CF) and (b) $pK^+\pi^-$ (DCS) combinations. DCS decay events are clearly observed in $M(pK^+\pi^-)$. In order to obtain the signal yield, a binned least-$\chi^2$ fit is performed. From the mass fit, we extract $(1.452\pm0.015)\times10^6$ $\Lambda_c^+\to pK^-\pi^+$ events and $3587\pm380$ $\Lambda_c^+\to pK^+\pi^-$ events. The latter has a peaking background from the single Cabibbo-suppressed (SCS) decay $\Lambda_c^+\to\Lambda(\to p\pi^-)K^+$, which has the same final-state topology. After subtracting the SCS contribution, we have $3379\pm380\pm78$ DCS events, where the first uncertainty is statistical and the second is the systematic due to SCS subtraction. The corresponding statistical significance is 9.4 standard deviations. We measure the branching ratio,
$\frac{\mathcal{B}(\Lambda_c^+\to pK^+\pi^-)}{\mathcal{B}(\Lambda_c^+\to pK^-\pi^+)}=(2.35\pm0.27\pm0.21)\times10^{-3}$, and the absolute branching fraction of the DCS decay, $\mathcal{B}(\Lambda_c^+\to pK^+\pi^-)=(1.61\pm0.23^{+0.07}_{-0.08})\times10^{-4}$. This measured branching ratio corresponds to $(0.82\pm0.21)\tan^4\theta_c$, where the uncertainty is the total, which suggests a slightly smaller decay width than the na{\"i}ve expectation~\cite{Link:2005ym}.
After subtracting the contributions of $\Lambda^*(1520)$ and $\Delta$ isobar intermediates, which contribute only to the CF decay, the revised ratio,  $\frac{\mathcal{B}(\Lambda_c^+\to pK^+\pi^-)}{\mathcal{B}(\Lambda_c^+\to pK^-\pi^+)}=(1.10\pm0.17)\tan^4\theta_c$ is consistent with the na{\"i}ve expectation.

%%%%%%%%%%%%%%%%%%%%%%%%%%%%%%%%%%%%%%%%%%%%%%%%%%%%%%
\subsection{Study of excited $\Xi_c$ states decaying into $\Xi_c^0$ and $\Xi_c^+$ baryons}
We present measurements of the masses of all members of the $\Xi_c'$, $\Xi_c(2645)$, $\Xi_c(2790)$, $\Xi_c(2815)$ and $\Xi_c(2980)$ isodoublets, measurements of the intrinsic widths of those that decay strongly, and evidence of previously unknown transitions~\cite{Yelton:2016fqw}. This analysis is based on $980~{\rm fb}^{-1}$ of data. The following decay chains are used: 
$\Xi_c(2980)\to\Xi_c(2645)\pi\to\Xi_c\pi\pi$, $\Xi_c(2980)\to\Xi_c'\pi\to\Xi_c\gamma\pi$, $\Xi_c(2815)\to\Xi_c(2645)\pi\to\Xi_c\pi\pi$, $\Xi_c(2815)\to\Xi_c'\pi\to\Xi_c\gamma\pi$ and $\Xi_c(2790)\to\Xi_c'\pi\to\Xi_c\gamma\pi$. To obtain large statistics, we use many decay modes of the ground-state $\Xi_c^0$ and $\Xi_c^+$ baryons.

Table~\ref{tab:xic} shows the results of the measurements of masses and widths of the five isodoublets.
\begin{table}[htb]
\renewcommand{\arraystretch}{1.2}
\caption{\small The results of the measurements of masses (in ${\rm MeV}/c^2$)  and widths (in MeV) of the five isodoublets. In all cases, the first
uncertainty is statistical, and the second is the systematic uncertainty associated with the individual measurement.
All the masses have a final, asymmetric uncertainty, taken from the Particle Data Group~\cite{Agashe:2014kda}, for the mass of the ground
states, and the $\Xi_c(2790)$ have an extra uncertainty due to the uncertainty in the $M(\Xi_c')-M(\Xi_c)$  measurement.}
\label{tab:xic}
\centering
\begin{tabular}{l|c|c}\hline
\hline
Particle & Mass & Width\\
\hline
$\Xi_c(2645)^+$ & $2645.58\pm0.06\pm0.07^{+0.28}_{-0.40}$ & $2.06\pm0.13\pm0.13$\\
$\Xi_c(2645)^0$ & $2646.43\pm0.07\pm0.07^{+0.28}_{-0.40}$ & $2.35\pm0.18\pm0.13$\\
\hline
$\Xi_c(2815)^+$ & $2816.73\pm0.08\pm0.06^{+0.28}_{-0.40}$ & $2.43\pm0.20\pm0.17$\\
$\Xi_c(2815)^0$  & $2820.20\pm0.08\pm0.07^{+0.28}_{-0.40}$ & $2.54\pm0.18\pm0.17$ \\
\hline
$\Xi_c(2980)^+$ & $2966.0\pm0.8\pm0.2^{+0.3}_{-0.4}$ & $28.1\pm2.4^{+1.0}_{-5.0}$\\
$\Xi_c(2980)^0$ &  $2970.8\pm0.7\pm0.2^{+0.3}_{-0.4}$ & $30.3\pm2.3^{+1.0}_{-1.8}$ \\
\hline
$\Xi_c'^+$ & $2578.4\pm0.1\pm0.4^{+0.3}_{-0.4}$ & \\
$\Xi_c'^0$ & $2579.2\pm0.1\pm0.4^{+0.3}_{-0.4}$ & \\
\hline
$\Xi_c(2790)^+$ & $2791.6\pm0.2\pm0.1\pm0.4^{+0.3}_{-0.4}$ & $8.9\pm0.6\pm0.8$\\
$\Xi_c(2790)^0$ &  $2794.9\pm0.3\pm0.1\pm0.4^{+0.3}_{-0.4}$ &$10.0\pm0.7\pm0.8$\\
\hline
\hline
\end{tabular}
\end{table}
Of the eighteen measurements, five are of intrinsic widths of particles for
which only limits existed previously. Of the remaining thirteen measurements, ten are within one standard deviation
of the Particle Data Group~\cite{Agashe:2014kda}  best-fit values. The three measurements that are in modest disagreement with previous
results are in the $\Xi_c(2980)$ sector, where the previous measurements were dominated by decays into different final
states and for which some measurements may have been prone to the existence of more than one resonance in the
region, or biases from threshold effects. Measurements of the isospin splittings, give in Table~\ref{tab:xic_iso} are in good agreement with predictions of non-relativistic quark model~\cite{SilvestreBrac:2003kd}.
\begin{table}[htb]
\caption{\small Isospin splitting between the members of each isodoublet.}
\label{tab:xic_iso}
\centering
\begin{tabular}{l|c}\hline
\hline
Particle & $M(\Xi_c^+)-M(\Xi_c^0)$ (${\rm MeV}/c^2$) \\
\hline
$\Xi_c(2645)$ & $-0.85\pm0.09\pm0.08\pm0.48$\\
$\Xi_c(2815)$ & $-3.47\pm0.12\pm0.05\pm0.48$\\
$\Xi_c(2980)$ & $-4.8\pm0.1\pm0.2\pm0.5$\\
$\Xi_c'$            &$-0.8\pm0.1\pm0.1\pm0.5$\\
$\Xi_c(2790)$  &$-3.3\pm0.4\pm0.1\pm0.5$\\
\hline
\hline
\end{tabular}
\end{table}
%%%%%%%%%%%%%%%%%%%%%%%%%%%%%%%%%%%%%%%%%%%%%%%%%%%%%%
\subsection{Studies of charmed strange baryons in the $\Lambda D$ final state}
To date, all measurements of the excited $\Xi_c$ were performed using decays in which the charm quark is contained in the final state baryon. Measurements of final states in which the charm quark is part of the final state meson provide complementary information. Here we present the 
studies of $\Xi^*_c$ baryons decaying to $\Lambda D^+$ and $\Lambda D^0$ final states using  $980~{\rm fb}^{-1}$ of data~\cite{Kato:2016hca}.

\begin{figure}[htb!]
\begin{center}
\includegraphics[width=\textwidth]{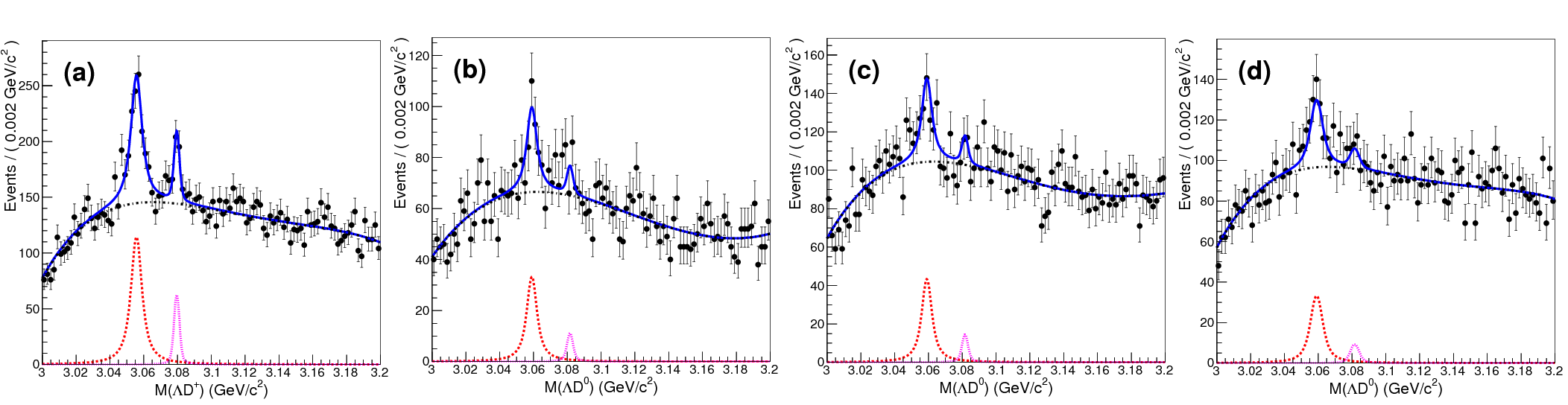}
\vskip -0.1cm
\caption{\small Distributions of (a) $M(\Lambda D^+)$ and (b, c, d) $M(\Lambda D^0)$  for (b) $D^0\to K^-\pi^+$, (c) $D^0\to K^-\pi^+\pi^+\pi^-$  and (d) $D^0\to K^-\pi^+\pi^0$. Points with statistical error bars are data. Blue solid lines show the fit results. The red dashed, magenta dotted, and black dashed-dotted lines show the $\Xi_c(3055)$ signal, the $\Xi_c(3080)$ signal, and the background components, respectively.}
\label{fig:lamd1}
\end{center}
\end{figure}

Figures~\ref{fig:lamd1} (a) show the fit to the $\Lambda D^+$ invariant mass distribution. Peaks of $\Xi_c(3055)^+$ and $\Xi_c(3080)^+$ states with statistical significances of 11.7 and 4.8 standard deviations, respectively are seen in the $\Lambda D^+$ final state.  These correspond to  first observation of  $\Xi_c(3055)^+\to\Lambda D^+$ and first evidence of $\Xi_c(3080)^+\to\Lambda D^+$ decays. The mass and width of the $\Xi_c(3055)^+$ are obtained to be $(3055.8\pm0.4\pm0.2)~{\rm MeV}/c^2$ and $(7.0\pm1.2\pm1.5) ~{\rm MeV}$, respectively, and those for $\Xi_c(3080)^+$ are $(3079.6\pm0.4\pm0.1)~{\rm MeV}/c^2$ and $<6.3$ MeV at 90\% confidence level, respectively. The measured  values for $\Xi_c(3055)^+$ are more accurate than the current world average values.
Figure~\ref{fig:lamd1} (b, c, d) shows the simultaneous fit to the three different $D^0$ decay modes of $\Lambda D^0$ final state. We observe a clear peak of $\Xi_c(3055)^0$ state with a statistical significance of 8.6 standard deviations. The peak for the $\Xi_c(3080)^0$ is not statistically significant. The mass and width of $\Xi_c(3055)^0$ are measured to be $(3059.0\pm0.5\pm0.6)~{\rm MeV}/c^2$ and $(6.4\pm2.1\pm1.1)~{\rm MeV}$, respectively. This is the first observation of the decay $\Xi_c(3055)^0\to\Lambda D^0$.

We perform a combined analysis of the  $\Xi_c(3055)^+$ and $\Xi_c(3080)^+$ by comparing their decays into $\Lambda D^+$ with those into $\Sigma_c^{++}K^-$ and $\Sigma^{*++}_cK^-$ final states. We measure the ratios of branching fractions $\frac{\mathcal{B}(\Xi_c(3055)^+\to\Lambda D^+)}{\mathcal{B}(\Xi_c(3055)^+\to\Sigma_c^{++}K^-)}=5.09\pm1.01\pm0.76$, $\frac{\mathcal{B}(\Xi_c(3080)^+\to\Lambda D^+)}{\mathcal{B}(\Xi_c(3080)^+\to\Sigma_c^{++}K^-)}=1.29\pm0.30\pm0.15$ and \newline$\frac{\mathcal{B}(\Xi_c(3080)^+\to\Sigma^{*++}K^-)}{\mathcal{B}(\Xi_c(3080)^+\to\Sigma_c^{++}K^-)}=1.07\pm0.27\pm0.04$. These measurements are in contradictions with the chiral quark model predictions~\cite{Liu:2012sj}. Further experimental and theoretical work is needed to understand the nature of these baryons.  From the combined analysis we measure the widths of the state $\Xi_c(3055)^+$ is $(7.8\pm1.2\pm1.5)$ MeV and that of $\Xi_c(3080)^+$ is $(3.0\pm0.7\pm0.4)$ MeV.
%%%%%%%%%%%%%%%%%%%%%%%%%%%%%%%%%%%%%%%%%%%%%%%%%%%%%%
%======================================================================
\section{Analysis of charmed meson decays}
\subsection{$D^0\to V \gamma$}
\begin{figure}[h!tb]
\begin{center}
\includegraphics[width=\textwidth]{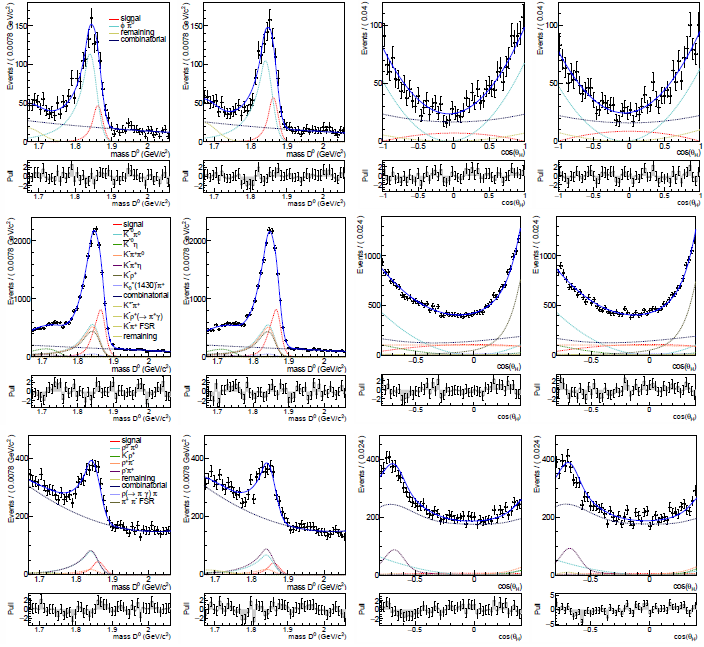}
\caption{\small Distributions of the  invariant masses and the cosine of the helicity angles of (left) $D^0\to V\gamma$ and (right) $\bar{D^0}\to V\gamma$. Top row corresponds to $\phi$ mode, middle row the $\bar{K}^{*0}$ and the bottom row the $\rho^0$ mode.}
\label{fig:d2vgam}
\end{center}
\end{figure}

The radiative decays $D^0\to V \gamma$, where $V$  $(\phi, \bar{K}^{*0}, \rho^0)$ is a vector meson, are dominated by  long-range contribution. They could be sensitive to NP via $CP$ asymmetry~\cite{Isidori:2012yx, Lyon:2012fk}. We present here the  results of the measurement of the branching fractions and
$CP$ asymmetries in these decays based on $943~{\rm fb}^{-1}$ of data~\cite{Abdesselam:2016yvr}. The candidate $D^0$ mesons are required to originate from the decay $D^{*+}\to D^0\pi^{+}_{S}$ in order to identify the flavour of the $D^0$ and to suppress combinatorial background.
The signal decays are reconstructed in the following sub-decay channels of the vector meson: $\phi\to K^+K^-$, $\bar{K}^{*0}\to K^-\pi^+$ and $\rho^0\to\pi^+\pi^-$. The selection criteria are optimized to maximize the figure of merit $N_{\rm sig}/\sqrt{N_{\rm sig}+N_{\rm bkg}}$, where $N_{\rm sig}$ and $N_{\rm bkg}$ represent the number of signal and background events. Both the branching fractions $\mathcal{B}(D^0\to V \gamma)$ and 
$CP$ asymmetries $\mathcal{A}_{CP}(D^0\to V \gamma)$ are obtained via normalization to other decay channels. The signal branching fraction
$\mathcal{B}_{\rm sig}$ is given by
\begin{equation}
\mathcal{B}_{\rm sig}=\mathcal{B}_{\rm norm}\times \frac{N_{\rm sig}}{N_{\rm norm}}\times\frac{\varepsilon_{\rm norm}}{\varepsilon_{\rm sig}},
\end{equation}
where $N$ is the extracted yield, $\varepsilon$ the reconstruction efficiency and $\mathcal{B}$ the branching fraction for the corresponding mode. For
$\mathcal{B}_{\rm norm}$ the world average value~\cite{Agashe:2014kda} is used. We use the decays $D^0\to K^+K^-$, $D^0\to K^-\pi^+$ and $D^0\to\pi^+\pi^-$ as the normalization channels for $\phi$, $\bar{K}^{*0}$ and $\rho^0$ signal modes, respectively. The measured branching fractions are
\begin{eqnarray*}
\mathcal{B}(D^0\to \phi \gamma) &=& (2.76\pm0.19\pm0.10)\times10^{-5},\\
\mathcal{B}(D^0\to  \bar{K}^{*0} \gamma) &=& (4.66\pm0.21\pm0.21)\times10^{-4},\\
\mathcal{B}(D^0\to \rho^0 \gamma) &=& (1.77\pm0.30\pm0.07)\times10^{-5}.
\end{eqnarray*}
The result of the $\phi$ mode is improved compared to the previous Belle result~\cite{Abe:2003yv} and is consistent with the world average value~\cite{Agashe:2014kda}. Our branching fraction for the $\bar{K}^{*0}$ mode is 3.3 $\sigma$ higher than the BaBar result $\mathcal{B}(D^0\to  \bar{K}^{*0} \gamma)=(3.22\pm0.20\pm0.27)\times10^{-4}$~\cite{Aubert:2008ai}. For the $\rho^0$ mode, we report the first observation of the decay with a significance of 5~$\sigma$ including the systematic uncertainties. The fit results are shown in Fig.~\ref{fig:d2vgam}.

The raw asymmetry is extracted using 
\begin{equation}
A_{\rm raw}=\frac{N(D^0)-N(\bar{D^0})}{N(D^0)+N(\bar{D^0})},
\end{equation}
and has several contributions: $A_{\rm raw}=\mathcal{A}_{CP}+A_{\rm FB}+A^{\pm}_{\varepsilon}$. The forward-backward production asymmetry $A_{\rm FB}$ and the asymmetry due to different reconstruction efficiencies for pasitively and negatively chraged particles $A^{\pm}_{\varepsilon}$  can be eliminated through a relative measurement of $\mathcal{A}_{CP}$, if the charged final state particles are identical. It then follows
\begin{equation}
\mathcal{A}_{CP}^{\rm sig}=A_{\rm raw}^{\rm sig}-A_{\rm raw}^{\rm norm}+\mathcal{A}_{CP}^{\rm norm},
\end{equation}
where $\mathcal{A}_{CP}^{\rm norm}$ is the nominal value of $CP$ asymmetry~\cite{Agashe:2014kda} of the normalization mode. We measure
\begin{eqnarray*}
\mathcal{A}_{CP}(D^0\to \phi \gamma) &=& -(0.094\pm0.066\pm0.001),\\
\mathcal{A}_{CP}(D^0\to  \bar{K}^{*0} \gamma) &=& -(0.003\pm0.020\pm0.000),\\
\mathcal{A}_{CP}(D^0\to \rho^0 \gamma) &=& ~~~~0.056\pm0.152\pm0.006.
\end{eqnarray*}
These are the first-ever $\mathcal{A}_{CP}$ measurements for these decays and are consistent with no $CP$ asymmetry in these modes.
%===================================================================================
\subsection{$D^0\to K_S^0K_S^0$}
In this analysis, we report the preliminary result of a measurement of  $CP$  asymmetry in $D^0\to K_S^0K_S^0$ decays, using $921~{\rm fb}^{-1}$ of data~\cite{Abdesselam:2016gqq}. The $CP$ asymmetry of the signal decay mode is given by
\begin{equation}
\mathcal{A}_{CP}^{\rm sig}=A_{\rm raw}^{\rm sig}-A_{\rm raw}^{\rm norm}+\mathcal{A}_{CP}^{\rm norm}+A^{K}_{\varepsilon}.
\end{equation}
The decay  $D^0\to K_S^0\pi^0$ is used as the normalization mode. Here $A^{K}_{\varepsilon}$ is the asymmetry due to strong interaction of $K^0$ and $\bar{K}^0$ mesons with nucleons of the detector material.
\begin{figure}[h!tb]
\begin{center}
\includegraphics[width=\textwidth]{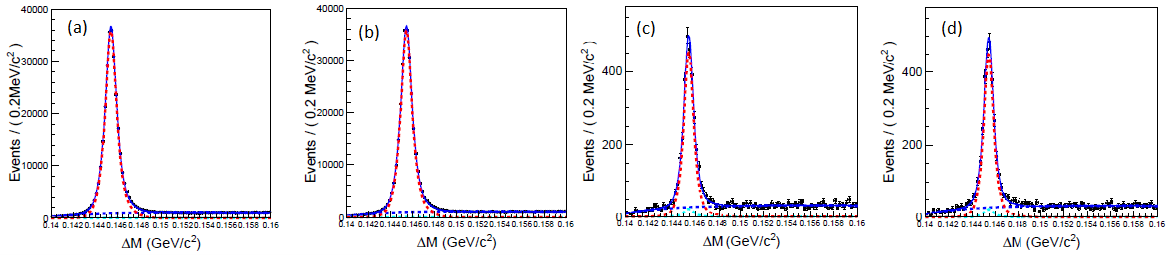}
\caption{\small Distributions of $\Delta M$ for the $K_S^0\pi^0$ [(a) and (b)] and $K^0_SK^0_S$ [(c) and (d)] final states. (a) and (c) for $D^{*+}$ decays and (b) and (d) for $D^{*-}$ decays. Here points with error bars are the data, the solid curves represent the total fit, dashed blue, cyan and red  curves show the non-peaking background,  peaking background and  signal, respectively.}
\label{fig:d2ksks}
\end{center}
\end{figure}
A simultaneous fit of the $\Delta M$ for $D^{*+}$ and $D^{*-}$ is used to measure the raw asymmetry, shown in Fig.~\ref{fig:d2ksks}. We measure
\begin{eqnarray*}
\mathcal{A}_{CP}(D^0\to K_S^0K_S^0)&=&-(0.02\pm1.53\pm0.17)\%. 
\end{eqnarray*}
The result is consistent with no $CP$ violation and with  SM expectations~\cite{Hiller:2012xm, Nierste:2015zra} and is a significant improvement compared to the previous measurements by CLEO~\cite{Bonvicini:2000qm} and LHCb~\cite{Aaij:2015fua}.

\section*{Acknowledgements}
The authors thank the workshop organizers for
hosting a fruitful and stimulating workshop and
providing excellent hospitality. This research is supported
by the U.S. Department of Energy.

%%%%%%%%%%%%%%%%%%%%%%%%%%%%%%%%%%%%%%%%%%


\begin{thebibliography}{99}

\bibitem{KEKB}
S.~Kurokawa and E.~Kikutani, 
%``Overview of the KEKB accelerators,"
Nucl. Instrum. Methods Phys. Res. Sect.
 A {\bf 499}, 1 (2003), and other papers included in this Volume;
 T.Abe {\it et al.}, 
%``Accelerator design at SuperKEKB,''
Prog. Theor. Exp. Phys. {\bf 2013}, 03A011 (2013)
 and references therein.

\bibitem{Belle}
A.~Abashian {\it et al.} (Belle Collaboration), 
%``The Belle Detector,''
Nucl. Instrum. Methods
 Phys. Res., Sect. A {\bf 479}, 117 (2002); also see the detector section in
 J.Brodzicka {\it et al.}, 
%``Physics achievements from the Belle Experiment,''
Prog. Theor. Exp. Phys. {\bf 2012}, 04D001 (2012). 

\bibitem{svd2}
  Z.~Natkaniec {\it et al.} (Belle SVD2 Group),
%``Status of the Belle silicon vertex detector,''
 Nucl. Instrum. Methods Phys. Res., Sect. A  {\bf 560}, 1 (2006).
%Y. Ushiroda (Belle SVD2 Group), Nucl. Instr. and Meth.A {\bf 511} 6 (2003). 

%\cite{Neubert:1993mb}
\bibitem{Neubert:1993mb} 
  M.~Neubert,
  %``Heavy quark symmetry,''
  Phys.\ Rept.\  {\bf 245}, 259 (1994).
  %doi:10.1016/0370-1573(94)90091-4
%  [hep-ph/9306320].
  %%CITATION = doi:10.1016/0370-1573(94)90091-4;%%
  %1328 citations counted in INSPIRE as of 27 Oct 2016


%\cite{Uraltsev:2000ce}
\bibitem{Uraltsev:2000ce} 
  N.~Uraltsev,
  %``New exact heavy quark sum rules,''
  Phys.\ Lett.\ B {\bf 501}, 86 (2001).
 % doi:10.1016/S0370-2693(01)00110-1
%  [hep-ph/0011124].
  %%CITATION = doi:10.1016/S0370-2693(01)00110-1;%%
  %94 citations counted in INSPIRE as of 27 Oct 2016

%\cite{Agashe:2014kda}
\bibitem{Agashe:2014kda} 
  K.~A.~Olive {\it et al.} (Particle Data Group),
  %``Review of Particle Physics,''
  Chin.\ Phys.\ C {\bf 38}, 090001 (2014).
 % doi:10.1088/1674-1137/38/9/090001
  %%CITATION = doi:10.1088/1674-1137/38/9/090001;%%
  %4863 citations counted in INSPIRE as of 26 Sep 2016

%\cite{Matvienko:2015gqa}
\bibitem{Matvienko:2015gqa} 
  D.~Matvienko {\it et al.} (Belle Collaboration),
  %``Study of $D^{**}$ production and light hadronic states in the $\bar{B}^0 \to D^{*+} \omega \pi^-$ decay,''
  Phys.\ Rev.\ D {\bf 92}, no. 1, 012013 (2015).
 % doi:10.1103/PhysRevD.92.012013
  %[arXiv:1505.03362 [hep-ex]].
  %%CITATION = doi:10.1103/PhysRevD.92.012013;%%
  %2 citations counted in INSPIRE as of 27 Oct 2016

%\cite{Alexander:2001fp}
\bibitem{Alexander:2001fp} 
  J.~P.~Alexander {\it et al.} (CLEO Collaboration),
  %``First observation of anti-B ---> D(*) rho-prime-, rho-prime- ---> omega pi-,''
  Phys.\ Rev.\ D {\bf 64}, 092001 (2001).
  %doi:10.1103/PhysRevD.64.092001
  %[hep-ex/0103021].
  %%CITATION = doi:10.1103/PhysRevD.64.092001;%%
  %27 citations counted in INSPIRE as of 27 Oct 2016

%\cite{Aubert:2006zb}
\bibitem{Aubert:2006zb} 
  B.~Aubert {\it et al.} (BaBar Collaboration),
  %``Study of the decay $\bar{B}^0 \to D^{*+} \omega \pi^{-}$,''
  Phys.\ Rev.\ D {\bf 74}, 012001 (2006).
  %doi:10.1103/PhysRevD.74.012001
  %[hep-ex/0604009].
  %%CITATION = doi:10.1103/PhysRevD.74.012001;%%
  %19 citations counted in INSPIRE as of 27 Oct 2016

%\cite{Matvienko:2011ic}
\bibitem{Matvienko:2011ic} 
  D.~V.~Matvienko, A.~S.~Kuzmin and S.~I.~Eidelman,
  %``A model of $\bar{B}^0\to D^{*+}\omega\pi^-$ decay,''
  JHEP {\bf 1109}, 129 (2011).
 % doi:10.1007/JHEP09(2011)129
  %[arXiv:1108.2862 [hep-ph]].
  %%CITATION = doi:10.1007/JHEP09(2011)129;%%
  %2 citations counted in INSPIRE as of 27 Oct 2016

%%%%%%%%%%%%%%%%%%%%%%%%%%%%%%%%%%%%%%%%%%
%\cite{Lipkin:2002za}
\bibitem{Lipkin:2002za} 
  H.~J.~Lipkin,
  %``Puzzles in hyperon, charm and beauty physics,''
  Nucl.\ Phys.\ Proc.\ Suppl.\  {\bf 115}, 117 (2003).
  %doi:10.1016/S0920-5632(02)01965-5
  %[hep-ph/0210166].
  %%CITATION = doi:10.1016/S0920-5632(02)01965-5;%%
  %9 citations counted in INSPIRE as of 04 Nov 2016

%\cite{Gao:2006nb}
\bibitem{Gao:2006nb} 
  D.~N.~Gao,
  %``Strong phases, asymmetries, and SU(3) symmetry breaking in D ---> K pi decays,''
  Phys.\ Lett.\ B {\bf 645}, 59 (2007).
 % doi:10.1016/j.physletb.2006.11.069
  %[hep-ph/0610389].
  %%CITATION = doi:10.1016/j.physletb.2006.11.069;%%
  %20 citations counted in INSPIRE as of 04 Nov 2016


%\cite{Yang:2015ytm}
\bibitem{Yang:2015ytm} 
  S.~B.~Yang {\it et al.} (Belle Collaboration),
  %``First Observation of Doubly Cabibbo-Suppressed Decay of a Charmed Baryon: $\Lambda^{+}_{c} \rightarrow p K^{+} \pi^{-}$,''
  Phys.\ Rev.\ Lett.\  {\bf 117}, 011801 (2016).
  %doi:10.1103/PhysRevLett.117.011801
 % [arXiv:1512.07366 [hep-ex]].
  %%CITATION = doi:10.1103/PhysRevLett.117.011801;%%
  %2 citations counted in INSPIRE as of 04 Nov 2016


%\cite{Link:2005ym}
\bibitem{Link:2005ym} 
  J.~M.~Link {\it et al.} (FOCUS Collaboration),
  %``Search for Lambda+(c) ---> p K+ pi- and D+(s) ---> K+ K+ pi- using genetic programming event selection,''
  Phys.\ Lett.\ B {\bf 624}, 166 (2005).
 % doi:10.1016/j.physletb.2005.08.032
%  [hep-ex/0507103].
  %%CITATION = doi:10.1016/j.physletb.2005.08.032;%%
  %11 citations counted in INSPIRE as of 04 Nov 2016

%%%%%%%%%%%%%%%%%%%%%%%%%%%%%%%%%%%%%%%%%%

%\cite{Yelton:2016fqw}
\bibitem{Yelton:2016fqw} 
  J.~Yelton {\it et al.} (Belle Collaboration),
  %``Study of Excited $\Xi_c$ States Decaying into $\Xi_c^0$ and $\Xi_c^+$ Baryons,''
  Phys.\ Rev.\ D {\bf 94}, 052011 (2016).
  %doi:10.1103/PhysRevD.94.052011
  %[arXiv:1607.07123 [hep-ex]].
  %%CITATION = doi:10.1103/PhysRevD.94.052011;%%
  %2 citations counted in INSPIRE as of 07 Nov 2016

%\cite{SilvestreBrac:2003kd}
\bibitem{SilvestreBrac:2003kd} 
  B.~Silvestre-Brac, F.~Brau and C.~Semay,
  %``Electromagnetic splitting for mesons and baryons using dressed constituent quarks,''
  J.\ Phys.\ G {\bf 29}, 2685 (2003).
  %doi:10.1088/0954-3899/29/12/002
 % [hep-ph/0302252].
  %%CITATION = doi:10.1088/0954-3899/29/12/002;%%
  %11 citations counted in INSPIRE as of 07 Nov 2016
%%%%%%%%%%%%%%%%%%%%%%%%%%%%%%%%%%%%%%%%%%


%\cite{Kato:2016hca}
\bibitem{Kato:2016hca} 
  Y.~Kato {\it et al.} (Belle Collaboration),
  %``Studies of charmed strange baryons in the $\Lambda$D final state at Belle,''
  Phys.\ Rev.\ D {\bf 94}, 032002 (2016).
  %doi:10.1103/PhysRevD.94.032002
 % [arXiv:1605.09103 [hep-ex]].
  %%CITATION = doi:10.1103/PhysRevD.94.032002;%%
  %5 citations counted in INSPIRE as of 04 Nov 2016

%\cite{Liu:2012sj}
\bibitem{Liu:2012sj} 
  L.~H.~Liu, L.~Y.~Xiao and X.~H.~Zhong,
  %``Charm-strange baryon strong decays in a chiral quark model,''
  Phys.\ Rev.\ D {\bf 86}, 034024 (2012).
  %doi:10.1103/PhysRevD.86.034024
  %[arXiv:1205.2943 [hep-ph]].
  %%CITATION = doi:10.1103/PhysRevD.86.034024;%%
  %14 citations counted in INSPIRE as of 07 Nov 2016
%%%%%%%%%%%%%%%%%%%%%%%%%%%%%%%%%%%%%%%%%%

%\cite{Isidori:2012yx}
\bibitem{Isidori:2012yx} 
  G.~Isidori and J.~F.~Kamenik,
  %``Shedding light on CP violation in the charm system via D to V gamma decays,''
  Phys.\ Rev.\ Lett.\  {\bf 109}, 171801 (2012).
  %doi:10.1103/PhysRevLett.109.171801
 % [arXiv:1205.3164 [hep-ph]].
  %%CITATION = doi:10.1103/PhysRevLett.109.171801;%%
  %61 citations counted in INSPIRE as of 04 Oct 2016

%\cite{Lyon:2012fk}
\bibitem{Lyon:2012fk} 
  J.~Lyon and R.~Zwicky,
  %``Anomalously large ${\cal O}_8$ and long-distance chirality from $A_{\rm CP}[D^0 \to (\rho^0,\omega) \gamma](t)$,''
  arXiv:1210.6546 [hep-ph].
  %%CITATION = ARXIV:1210.6546;%%
  %22 citations counted in INSPIRE as of 04 Oct 2016


%\cite{Abdesselam:2016yvr}
\bibitem{Abdesselam:2016yvr} 
  T.~ Nanut {\it et al.} (Belle Collaboration),
  %``Measurement of the branching fraction and $CP$ asymmetry in radiative $D^0 \to V \gamma$ decays,''
  arXiv:1603.03257 [hep-ex].
  %%CITATION = ARXIV:1603.03257;%%
  %1 citations counted in INSPIRE as of 26 Sep 2016


%\cite{Abe:2003yv}
\bibitem{Abe:2003yv} 
  K.~Abe {\it et al.} (Belle Collaboration),
  %``Observation of radiative decay D0 ---> phi gamma,''
  Phys.\ Rev.\ Lett.\  {\bf 92}, 101803 (2004).
  %doi:10.1103/PhysRevLett.92.101803
%  [hep-ex/0308037].
  %%CITATION = doi:10.1103/PhysRevLett.92.101803;%%
  %23 citations counted in INSPIRE as of 26 Sep 2016

%\cite{Aubert:2008ai}
\bibitem{Aubert:2008ai} 
  B.~Aubert {\it et al.} (BaBar Collaboration),
  %``Measurement of the Branching Fractions of the Radiative Charm Decays D0 ---> anti-K*0 gamma and D0 ---> phi gamma,''
  Phys.\ Rev.\ D {\bf 78}, 071101 (2008).
 % doi:10.1103/PhysRevD.78.071101
 % [arXiv:0808.1838 [hep-ex]].
  %%CITATION = doi:10.1103/PhysRevD.78.071101;%%
  %5 citations counted in INSPIRE as of 26 Sep 2016

%\cite{Abdesselam:2016gqq}
\bibitem{Abdesselam:2016gqq} 
  A.~Abdesselam {\it et al.} (Belle Collaboration),
  %``Measurement of $CP$ asymmetry in the $D^{0} \to K^0_S K^0_S$ decay at Belle,''
  arXiv:1609.06393 [hep-ex].
  %%CITATION = ARXIV:1609.06393;%%

%\cite{Hiller:2012xm}
\bibitem{Hiller:2012xm} 
  G.~Hiller, M.~Jung and S.~Schacht,
  %``SU(3)-flavor anatomy of nonleptonic charm decays,''
  Phys.\ Rev.\ D {\bf 87}, 014024 (2013).
 % doi:10.1103/PhysRevD.87.014024
  %[arXiv:1211.3734 [hep-ph]].
  %%CITATION = doi:10.1103/PhysRevD.87.014024;%%
  %35 citations counted in INSPIRE as of 04 Oct 2016

%\cite{Nierste:2015zra}
\bibitem{Nierste:2015zra} 
  U.~Nierste and S.~Schacht,
  %``CP Violation in $D^0\rightarrow K_SK_S$,''
  Phys.\ Rev.\ D {\bf 92},  054036 (2015).
  %doi:10.1103/PhysRevD.92.054036
%  [arXiv:1508.00074 [hep-ph]].
  %%CITATION = doi:10.1103/PhysRevD.92.054036;%%
  %2 citations counted in INSPIRE as of 04 Oct 2016

%\cite{Bonvicini:2000qm}
\bibitem{Bonvicini:2000qm} 
  G.~Bonvicini {\it et al.} (CLEO Collaboration),
  %``Search for CP violation in $D^0 \to K^0_{S} \pi^0$ and $D^0 \to \pi^0 \pi^0$ and $D^0 \to K^0_{S} K^0_{S}$ decays,''
  Phys.\ Rev.\ D {\bf 63}, 071101 (2001).
  %doi:10.1103/PhysRevD.63.071101
 % [hep-ex/0012054].
  %%CITATION = doi:10.1103/PhysRevD.63.071101;%%
  %43 citations counted in INSPIRE as of 04 Oct 2016

%\cite{Aaij:2015fua}
\bibitem{Aaij:2015fua} 
  R.~Aaij {\it et al.} (LHCb Collaboration),
  %``Measurement of the time-integrated $CP$ asymmetry in $D^0 \to K^0_S K^0_S$ decays,''
  JHEP {\bf 1510}, 055 (2015).
 % doi:10.1007/JHEP10(2015)055
%  [arXiv:1508.06087 [hep-ex]].
  %%CITATION = doi:10.1007/JHEP10(2015)055;%%
  %3 citations counted in INSPIRE as of 04 Oct 2016


\end{thebibliography}
\end{document}